\documentclass[conference,10pt]{IEEEtran}
\usepackage{ifpdf}
\ifpdf
\else
 \fi

\usepackage{cite}

\ifCLASSINFOpdf
\else
\fi
\usepackage[cmex10]{amsmath}
\usepackage{algorithmic}
\usepackage{algorithm}
\renewcommand{\baselinestretch}{0.99}

\usepackage{array}

\usepackage[tight,footnotesize]{subfigure}

\usepackage[utf8]{inputenc}
\usepackage[english]{babel}

\usepackage{amsthm}
\usepackage{mathtools}
\usepackage{graphicx}
\usepackage{geometry}
\newcommand{\subparagraph}{}
\usepackage[compact]{titlesec}
\titlespacing{\section}{0pt}{1ex}{1ex}
\titlespacing{\subsection}{0pt}{1ex}{1ex}
\usepackage[font=small,skip=2pt]{caption}

\begin{document}
%
\title{Sensing Throughput Optimization in Fading Cognitive Multiple Access Channels With Energy Harvesting Secondary Transmitters}

\author{\IEEEauthorblockN{Sinchan Biswas}
\IEEEauthorblockA{Signals and Systems Division\\
Uppsala University\\
Email: sinchan.biswas@angstom.uu.se}
\and
\IEEEauthorblockN{Amirpasha Shirazinia}
\IEEEauthorblockA{Signals and Systems Division\\
Uppsala University\\
Email:amirpasha.shirazinia@angstom.uu.se }
\and
\IEEEauthorblockA{Subhrakanti Dey\\
Signals and Systems Division\\
Uppsala University\\
Email:subhrakanti.dey@angstom.uu.se}}


%


\maketitle

\begin{abstract}
The paper investigates the problem of maximizing expected sum throughput in a fading multiple access cognitive radio  network when secondary user (SU) transmitters have energy harvesting capability, and perform cooperative spectrum sensing. We formulate the problem as maximization of sum-capacity of the cognitive multiple access network over a finite time horizon subject to a time averaged interference constraint at the primary user (PU)  and almost sure energy causality constraints at the SUs. The problem is a mixed integer non-linear program with respect to two decision variables  namely spectrum access decision and spectrum sensing decision, and the continuous variables sensing time and transmission power.  In general, this problem is known to be NP hard. For optimization over these two decision variables, we use an exhaustive search policy when the length of the time horizon is small, and  a heuristic policy for longer horizons. For given values of the decision variables, the problem simplifies into a joint optimization on SU \textit{transmission power} and \textit{sensing time}, which is non-convex in nature. We solve the resulting optimization problem as an alternating convex optimization problem for both non-causal  and causal channel state information and harvested energy information patterns at the SU base station (SBS) or fusion center (FC).   We present an analytic solution for the non-causal scenario with infinite battery capacity for a general finite horizon problem.We formulate the problem with causal information  and finite battery capacity as a stochastic control problem and solve it using the technique of dynamic programming. Numerical results are presented to illustrate the performance of the various algorithms.
\end{abstract}


%

\section{Introduction}

Spectrum scarcity is a significant issue in modern wireless networks. This is due to the legacy of static allocation policy of the radio spectrum, which prohibits  unlicensed users to exploit  licensed spectrum even when it is idle. As a solution, the paradigm of cognitive radio (CR) \cite{ref1} has been proposed. In the {\em interweave} paradigm of CR, based on the concept of dynamic spectrum allocation strategy,   unlicensed SUs can access the PU licensed spectrum when the PU is idle. The SUs have to vacate the licensed spectrum as soon as the PU becomes active. To achieve this, the SUs sense the spectrum to check whether the PU is active or not. The decision about spectrum sensing is then taken in a cooperative manner by sending all local spectrum sensing decision to an FC which makes an overall decision regarding spectrum access.
\par
The issue of energy efficiency is also a very important aspect of wireless transmission. While traditionally mobile devices have relied on rechargeable batteries, in many situations, periodic  battery replacement of the wireless nodes is not a feasible option in practice, such as in sensor networks. Thus energy harvesting from  natural sources like wind or solar power is a viable and cost-effective solution for replenishing energy. Recently, there has been significant research in the domain of energy harvesting in wireless environment \cite{ref2}. To analyze the performance of energy harvesting cognitive multiple access networks, throughput is generally used as the performance measure. For such a network, sum capacity with different wireless channel models \cite{ref3} has been investigated. As spectrum sensing is one of the key tasks in a CR network, sensing-throughput analysis has also been studied rigorously in \cite{ref4}. Capacity analysis of wireless system with energy harvesting capability has been studied in \cite{ref5} as well. Incorporating the capability of energy harvesting in CR network has also been investigated in \cite{ref6,ref7}. The research in this specific field focuses on both aspects of mitigating spectrum scarcity and efficient energy usage. Recently in \cite{ref8}, the authors have investigated  achievable throughput optimization in energy harvesting CR networks where the optimization is over sensing time and sensing threshold.
\par
As opposed to previous work in this area, we investigate a sensing-throughput optimization problem of a block fading multiple access CR network with a single PU, where all CR nodes (or SUs) are equipped with energy harvesting capability, and perform cooperative spectrum sensing. In this work, we investigate the tradeoff between sensing time and sum capacity of the CR multiple access channel with respect to \textit{transmission power} and \textit{sensing time}, keeping the sensing threshold fixed. Since spectrum sensing also consumes energy, due to the unreliable nature of harvested energy patterns, the CR nodes 
must make an initial decision as to whether they would perform spectrum sensing or not. All participating CR nodes' sensing decisions are combined at the SBS (also performing as the FC) to arrive at a final spectrum access decision which is broadcast to all CR nodes. If the PU is deemed to be absent, the participating CR nodes use the remaining time in each fading block for information transmission with a suitable transmission power. The problem we investigate is mixed integer nonlinear programming (MINLP) problem with respect to the individual spectrum sensing (whether or not to participate) and the overall spectrum access decisions, sensing time and transmission power of each user. We consider both non-causal and causal channel state information (CSI) and harvested energy information (or information on the battery level at each user) at the FC/SBS, for optimizing the CR multiple access channel sum capacity over a finite time horizon, where each user's energy consumption in each fading block cannot exceed their battery level at the beginning of the block (energy causality constraint), a peak transmission power constraint and an average interference constraint at the PU receiver. 
\par
Our contributions are listed as follows:
\begin{itemize}

\item The problem of sum-capacity maximization in a multiple access cognitive network environment is investigated with energy harvesting and both infinite/finite battery settings. We first derive an analytic closed form solution for the general horizon problem for the case of  non-causal CSI. Since the problem is individually convex with respect to the sensing time or the transmission power, but not jointly, we implement an iterative convex optimization scheme which is provably convergent to a locally optimal solution. 
\item For short horizon lengths, we employ an exhaustive search to find the optimal values of the Boolean variables, namely the spectrums sensing and access decisions. For longer horizons, motivated by the NP hard nature of the problem, we propose a heuristic algorithm to determine these variables in a suboptimal manner.
\item The problem involving the more realistic scenario of finite battery and causal CSI and harvested energy  is solved using the dynamic programming (DP) algorithm with discretized levels of power and sensing times.
\item We present illustrative numerical results to demonstrate the comparative performance of the various algorithms proposed in this work. 
\end{itemize}

\par

The rest of the paper is organized as follows. In section II we discuss the system model. In section III we describe the optimization problem. In section IV we propose the closed form solution of the problem for the infinite battery and non-causal CSI scenario. In section V we discuss the policy of causal finite battery scenario using the DP algorithm. Section VI contains simulation results, followed by concluding remarks  in section VII.

\begin{figure} [!ht]
  \centering
    \includegraphics[width=0.90\columnwidth, height=5cm ]{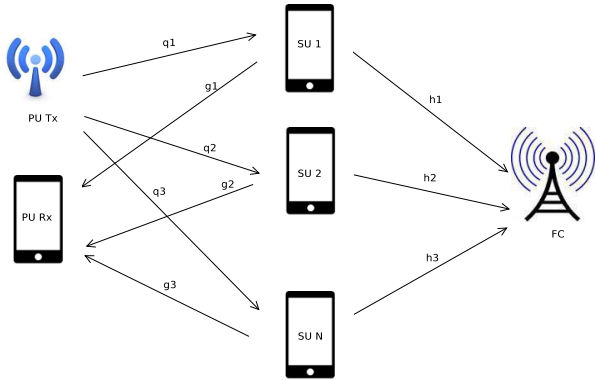}
  \caption{System model for Cognitive MAC.}\label{fig:diagram}
  \centering
\end{figure}

\section{System Model and Problem Formulation}

We study the system model as depicted in Fig.\ref{fig:diagram}. The system under consideration has $N$ number of SUs communicating to the SBS/FC.  In the adopted cooperative spectrum sensing model each SU decides whether to participate or not in sensing the PU spectrum (due to the random nature of harvested energy),  and sends its individual spectrum sensing decision (if participating) to the FC, which makes the overall decision and broadcasts it to all the SUs.
\par
We assume that  time is slotted, where each slot represents a block where the all fading channels remain invariant and change from one block to another in a statistically independent manner. All relevant parameters, random variables and optimization variables used in this paper are described in Table  \ref{sys_par}.  Each time slot of $T$ time units is utilized by the SUs for the spectrum sensing and data transmission. In  the $k^{th}$ time slot SUs spend $\tau_{k}$ time units for spectrum sensing and remainder of the slot $(T-\tau_{k})$ time units for data transmission, provided the spectrum sensing result indicates  PU spectrum vacancy. In the $k^{th}$ time slot, $i^{th}$ SU first makes a decision to perform spectrum sensing or to be idle on the basis of the individual decision to sense $a_{i,k}$, $1\leq i \leq N, 1 \leq k \leq M$, where $M$ is the length of time horizon over which the system performance is optimized, and $a_{i,k} \in \left\{0,1\right\}$, 1 (or 0) representing the decision to perform (or not to perform) spectrum sensing. The decision to sense $a_{i,k}$ is obtained by the individual SUs by the following rule \cite{ref9}
$a_{i,k}=1, \text{if} B_{i,k} \geq p_s \tau_{k}$ and $0$ otherwise,
where $p_s$ is the sensing power and $ B_{i,k}$ is the battery state in the $i^{th}$ SU at the beginning of the $k^{th}$ slot. With a battery of finite capacity $B_{max}$, $B_{i,k}$ can be expressed as:
\begin{equation} \label{eq:a2}
B_{i,k+1}=\min \left\{B_{max},B_{i,k}-E_{i,k}+H_{i,k}\right\}
\end{equation}
where  $H_{i,k}$ is the energy harvested by the $i^{th}$ SU in the $k^{th}$ time slot. In (\ref{eq:a2}), $E_{i,k}$ is the amount of energy used by $i^{th}$ SU at $k^{th}$ time slot, which  can be expressed as
\begin{equation}\label{eq:a3}
E_{i,k}=a_{i,k}(p_s\tau_{k}+p_{i,k}(T-\tau_{k})(1-\theta_{k}))
\end{equation}
where $\theta_k$ is the overall spectrum access decision indicating PU is present if $\theta_k = 1$ (details in the next subsection). 
One can also express  $B_{i,k}$ as
\begin{equation}\label{eq:a4}
B_{i,k}=\min \left\{B_{max},B_{i}-\sum_{r=1}^{k}E_{i,r}+ \sum_{r=1}^{k-1} H_{i,r}\right\} 
\end{equation}
where $B_{i}$ is the initial battery level of $i^{th}$ SU. Note that in the infinite battery scenario $B_{i,k}$ from (\ref{eq:a4}) simplifies to
$B_{i,k}=B_{i}-\sum_{r=1}^{k}E_{i,r}+ \sum_{r=1}^{k-1} H_{i,r}$.
\subsection{Spectrum Sensing Model}

If an SU decides to sense, it collects the samples of the received signal from PU by dividing sensing time interval $\tau_{k}$ to a number of mini-slots, where the length of the mini-slots is a constant and pre-decided. The PU spectrum availability is decided by the following received signal model under hypothesis $\mathcal{H}_{0}$ (PU absent) and 
$\mathcal{H}_{1}$ (PU present).
 \begin{eqnarray}\label{eq:a6}
 & & \mathcal{H}_{0}: y_{i,k,m}=n_{i,k,m} \nonumber \\
 & & \mathcal{H}_{1}: y_{i,k,m}=q_{i}x_{k}+n_{i,k,m} 
 \end{eqnarray}
 where $x_{k}$ denotes the PU's transmitted signal for the $k^{th}$ time slot, where it is assumed to be real valued and distributed as $x_{k} \sim \mathcal{N}(0,\sigma_{x}^{2})$. The parameter $q_{i}$ is the CSI between PU transmitter and $i^{th}$ SU sensing device, which is assumed to be known at the SU  throughout the spectrum sensing process. The parameters $y_{i,k,m}$ and $n_{i,k,m}$ are the real valued received signal and  noise components respectively for the $i^{th}$ SU, in $k^{th}$ time slot  and $m^{th}$ mini-slot. The noise is distributed as $n_{i,k,m} \sim \mathcal{N}(0,\sigma_{n}^{2})$. We adopt the energy detection policy of \cite{ref10} for each SU. The local spectrum sensing decision at the $i^{th}$ SU in $k^{th}$ horizon is defined as 
 $\theta_{i,k}=\mathcal{I} \left\{\frac{1}{S}\sum_{m=1}^{S}y_{i,k,m}^{2} \geq \gamma \right\} $,
 where $\mathcal{I}$ is the indicator function, $S$ is the number of mini-slots in a particular time slot and $\gamma$ is the detection energy threshold. These local decisions are sent to the FC by error-free control channels and combined using the OR logic fusion.
In this scenario, probability of false alarm is
$P_{fa}=Pr \left\{\theta_{k}=1|\mathcal{H}_{0}\right\}$.

We make the following implicit assumptions regarding  the system model under consideration. We consider the channel power gains between the SU transmitters and FC as well as  SU transmitters and PU receiver being distributed as exponential random variables with unity mean, without loss of generality.   It is also assumed that SUs have the ability of mitigating the interference caused by PU. This assumption simplifies the expression of sum capacity with respect to the scenario which accounts for PU interference as a random parameter, although the proposed algorithms in this paper can be extended to the case where PU interference can be explicitly considered.  We also model the primary user activity as a stationary random process, which is assumed to occur with probability $\kappa$  (of PU being present) within  each time slot.

\begin{table}[ht]
\small
\caption{System Parameters} 
\begin{tabular}{|l|p{7cm}|}
  \hline
  $\tau_{k}$ & Time taken to perform the spectrum 
   sensing in the $k^{th}$ slot.  \\
  $p_{i,k}$  &  Transmission power for 
   $i^{th}$  SU in the $k^{th}$ time slot. \\
   $\kappa$ & PU activity probability. \\
  $h_{i,k}$ & The CSI between $i^{th}$ SU Transmitter 
   and FC in the $k^{th}$ time slot.  \\
  $g_{i,k}$ & The CSI between $i^{th}$ SU Transmitter
  and PU Receiver in the $k^{th}$ time slot. \\
  $q_{i}$ & The CSI between PU Transmitter and
  and $i^{th}$ SU in the $k^{th}$ time slot. \\
  $p_s$ & Power allocated to sensing. \\
  $P_{max}$ & The peak power limit on $p_{i,k}$.  \\
  $B_{i}$ & The initial battery state for $i^{th}$ user.  \\
  $H_{i,k}$ & Energy harvested for $i^{th}$ SU for  
  the $k^{th}$ time slot. \\
  $a_{i,k}$ & Spectrum access decision variable for
   the $i^{th}$ SU in $k^{th}$ time slot. \\
  $\theta_{k}$ & Spectrum sensing decision for the $k^{th}$ horizon. \\
  $Q_{avg}$ & The average interference limit to the PU. \\
  \hline
\end{tabular}
\label{sys_par}
\end{table}
\vspace{7mm}

\subsection{Sum Capacity Maximization}
For a finite horizon of length $M$, the sensing-throughput optimization problem for a CR multiple access channel with an average interference constraint at the PU receiver and energy causality constraints at each SU node can be formulated as (with $\forall i, 1 \leq i \leq N, \forall k, 1 \leq k \leq M$)
{\small
\setlength{\arraycolsep}{-0.1em}
\begin{flalign}
&  \underset{a_{i,k},\theta_{k},p_{i,k},\tau_{k}}{\text{max}}
 \mathrm{E}\left\{\sum_{k=1}^{M}\frac{T-\tau_{k}}{MT}\log _{2}(1+\sum_{i=1}^{N}p_{i,k}h_{i,k}a_{i,k}(1-\theta_{k}))\right\} \label{eq:e9}\\
  \text{s.t.}
 & \frac{1}{M}\sum_{k=1}^{M}\mathrm{E} \left\{\frac{T-\tau_{k}}{T}\sum_{i=1}^{N}p_{i,k}g_{i,k}a_{i,k}(1-\theta_{k})\right\} \leq Q_{avg} \label{eq:e10} \\
 & 0 \leq p_{i,k} \leq P_{max};  \label{eq:e1}  \\
 & 0  \leq \tau_{k} \leq T;  \label{eq:e2} \\
& E_{i,k} \leq B_{i,k} \hspace{2mm} \textbf{a.s.} \label{eq:e3} 
\end{flalign}}
where the last constraint can be replaced by 
$  \sum_{r=1}^{k}a_{i,r}(p_s\tau_{r}+p_{i,r}(T-\tau_{r})(1-\theta_{r}) \leq B_{i} 
 +\sum_{r=1}^{k-1}H_{i,r}$ in the case of infinite battery capacity, 
and  \textbf{a.s.} stands for {\em almost surely}. Note also the peak transmission power constraint on each SU, motivated by practical scenarios.
\setlength{\arraycolsep}{3pt}

\textbf{Remark} It is important to note that the average interference term in the constraint (\ref{eq:e10}) is normalized by the primary activity factor $\kappa$, where $Q_{avg}=Q/{\kappa}$, $Q$ being the actual interference limit, since no interference is caused when the primary is not active. Note also that although the optimization variables indicate 
$a_{i,k},\theta_{k}, \tau_{k}$ separately, they are interdependent due to the dependence of  $a_{i,k}$ and $\tau_k$, and the dependence of the decision variables $\theta_{i,k}$ on $\tau_k$. 

The above mentioned optimization problem is a MINLP problem with respect to the allocated transmission power $p_{i,k}$, the sensing time $\tau_{k}$, decision to sense $a_{i,k}$ and spectrum sensing result $\theta_{k}$. In principle, one can employ computationally intensive global optimization methods (e.g. see \cite{ref11}) to solve such NP hard problems. In this paper, for short horizon lengths, we employ an exhaustive search policy for determining the integer variables $a_{i,k},  \theta_k$. For longer horizons, we will employ a heuristic policy to be described in the next subsection.  For the exhaustive search policy, the optimization is performed at the SBS, and all relevant decision variables are exchanged between the SUs and SBS via control channels. 


It should be noted that for a fixed choice of  $a_{i,k}$ and $\theta_{k}$ , the problem becomes an optimization over $p_{i,k}$ and $\tau_{k}$, which is a non-convex problem in general jointly in $p_{i,k}$ and $\tau_{k}$.
However, the problem becomes convex if we fix any one of the variables $\tau_{k}$ or $p_{i,k}, \forall i$, which can be solved by convex optimization algorithms. To solve this problem we take the approach of alternating  convex optimization (see e.g. \cite{ref12}), where one alternates between optimizing over $\tau_k$ with fixed $p_{i,k}$ and vice versa until the algorithm converges to a local minimum.

\subsection{Heuristic Policy for $a_{i,k}$ and $\theta_{k}$}
The exponential complexity of an exhaustive search algorithm for optimizing over $a_{i,k}$ and $\theta_{k}$ can quickly explode for large values  of $M$ and $N$. For this scenario, we propose  a heuristic suboptimal policy as described below. This heuristic policy is based on the idea that limiting the maximum value of  $P_{fa}$ on individual SUs would in turn impose a lower bound $\tau_{l}$ on \textit{sensing time} for each horizon. If $P_{fa}$ is bounded by the constraint $P_{fa} \leq \alpha$, for some $\alpha>0$, then we can have a corresponding lower bound for $\tau_{k}$ as $\tau_{l} \leq \tau_{k}$, which can be  obtained by  \cite{ref4}
$ \tau_{l}=\frac{1}{f_{s}} \left\{\frac{\mathbf{Q}^{-1}(\alpha)}{\frac{\gamma}{\sigma^{2}_{n}} -1}\right\}^{2}$.  
where $\mathbf{Q}^{-1}(\cdot)$ is the inverse complementary distribution function of the standard Gaussian random variable. 
Then $a_{i,k}=1$ if $B_{i,k} >  p_s \tau_l$ and vice versa. $\theta_{i,k}$ is determined on the basis of the particular $\tau_{k}$ resulting from the alternating convex optimization method, and the final decision $\theta_{k}$ is obtained by
combining the individual  $\theta_{i,k}$ at the FC  using the OR fusion rule.

\section{Non-Causal Optimization with Infinite Battery}
In this section we analyze the problem first in the context of non-causal CSI ($g_{i,k},h_{i,k}$) and harvested energy  ($H_{i,k}$)  scenario with the assumption of infinite battery, for fixed values of the integer variables $a_{i,k}$ and $\theta_k$ (obtained either via exhaustive search or the heuristic policy). We incorporate the aforementioned alternating convex optimization scheme. For the general $M$-horizon problem, we obtain a closed form solution for the optimization problem involving $p_{i,k}$, when $\tau_k$ is fixed. We also show that the optimization involving $\tau_{k}$ (when $p_{i,k}$ is fixed) is a linear programming problem, which can be solved by any established LP solver.

\subsection{Optimal Power Allocation Policy}
For a fixed $\tau_{k}$, the optimization problem for \textit{transmit power} is given below:
\begingroup
\setlength\abovedisplayskip{1pt}
\setlength\belowdisplayskip{1pt}
\setlength{\arraycolsep}{-0.1em}
\begin{align}
& \underset{p_{i,k}}{\text{max}}
& \sum_{k=1}^{M}\frac{T-\tau_{k}}{MT \log{2}}\ln(1+\sum_{i=1}^{N}p_{i,k}h_{i,k}a_{i,k}(1-\theta_{k}))  \nonumber\\
& \text{s.t.}
&  \frac{1}{M}\sum_{k=1}^{M}\frac{T-\tau_{k}}{T}\sum_{i=1}^{N}p_{i,k}g_{i,k}a_{i,k}(1-\theta_{k})\leq Q_{avg} \\
& & 0 \leq p_{i,k} \leq P_{max}; 1 \leq i \leq N, 1 \leq k \leq M  \label{eq:solve}\\
& & \sum_{r=1}^{k}a_{i,r}(p_s\tau_{r}+p_{i,r}(T-\tau_{r})(1-\theta_{r})) \leq B_{i}+\sum_{r=1}^{k-1}H_{i,r} \label{eq:sol}
\end{align}
\endgroup

\textbf{Remark}: For the above problem the expectations  with respect to (\ref{eq:e9}) and (\ref{eq:e10}) are removed from the objective function because the random variables corresponding to the channel gains and energy arrival process are non-causally known.
\par
The Lagrangian of the above mentioned convex optimization problem is given by
\begingroup
\setlength\abovedisplayskip{1pt}
\setlength\belowdisplayskip{1pt}
\setlength{\arraycolsep}{-0.1em}
\begin{flalign}
& \mathcal{L}(\left\{p_{i,k}\right\},\lambda,\delta_{i,k},\eta_{i,k},\mu_{i,k}) = \nonumber \\
& \sum_{k=1}^{M}\frac{T-\tau_{k}}{MT \log{2}} \times 
 \ln (1+\sum_{i=1}^{N}p_{i,k}h_{i,k}a_{i,k}(1-\theta_{k})) \nonumber\\
& - \lambda(\frac{1}{M}\sum_{k=1}^{M}\frac{T-\tau_{k}}{T}\sum_{i=1}^{N}p_{i,k}g_{i,k}a_{i,k}(1-\theta_{k})-Q_{avg}) \nonumber\\
& +\sum_{k=1}^{M}\sum_{i=1}^{N}\eta_{i,k}p_{i,k}-\sum_{k=1}^{M}\sum_{i=1}^{N}\delta_{i,k}(p_{i,k}-P_{max}) \nonumber \\
& -\sum_{k=1}^{M}\sum_{i=1}^{N}\mu_{i,k}(\sum_{r=1}^{k}a_{i,r}(p_s\tau_{r}+p_{i,r} (T-\tau_{r})(1-\theta_{j})) \nonumber \\
& \;\;\;\;\; \;\;\;\;\;\;\;\;\;\;\;\;\;\;\; -B_{i}-\sum_{r=1}^{k-1}H_{i,r} ) 
\end{flalign}
\endgroup
where $\lambda,\eta_{i,k},\delta_{i,k}$ and $\mu_{i,k}$ are the non-negative dual variables associated with the average interference constraint, transmission power lower bound, upper bound and the energy causality constraint respectively. Now we define $\mathcal{D}$ as the set of $p_{i,k}$'s which satisfies (\ref{eq:solve}) and (\ref{eq:sol}). The Lagrange dual function is defined as:
$g(\lambda)=\max_{\left\{p_{i,k}\right\} \in \mathcal{D}} \mathcal{L}(\left\{p_{i,k}\right\},\lambda,\eta_{i,k},\delta_{i,k},\mu_{i,k}) $,
and the dual problem is defined as
 $\min_{\lambda \geq 0} g(\lambda)$.
 The Lagrange dual function $g(\lambda)$ can be obtained by solving the corresponding optimization problem using the following Karush-Kuhn-Tucker (KKT) conditions, where the optimal primal and dual solutions are denoted as $p_{i,k}^{\star}$, $\mu_{i,k}^{\star}$, $\delta_{i,k}^{\star}$, $\eta_{i,k}^{\star}$: 
\begin{flalign} \label{eq:eq1}
& \frac{T-\tau_{k}}{MT\log{2}}\frac{h_{i,k}a_{i,k}(1-\theta_{k})}{1+\sum_{l=1}^{N}p_{i,k}^{\star}h_{i,k}a_{i,k}(1-\theta_{k})}+\eta_{i,k}^{\star}-\delta_{i,k}^{\star} \nonumber \\
& -\frac{T-\tau_{k}}{MT} \lambda g_{i,k}a_{i,k}(1-\theta_{k}) \nonumber \\
&- \sum_{r=k}^{M}a_{i,r}(T-\tau_{r})(1-\theta_{r})\mu_{i,r}^{\star}=0 \;  \forall i,k  \nonumber\\
& \eta_{i,k}^{\star}p_{i,k}^{\star}=0 \; \forall i,k \nonumber \\
& \delta_{i,k}^{\star}(p_{i,k}^{\star}-P_{max})=0 \;\forall i,k \nonumber \\
& p_{i,k}^{\star} \leq P_{max} \; \forall i,k 
\end{flalign}
with $p_{i,k}^{\star} \geq 0,\delta_{i,k}^{\star} \geq 0, \eta_{i,k}^{\star} \geq 0, \mu_{i,k}^{\star} \geq 0$. Also $p_{i,k}^{\star}$, $\mu_{i,k}^{\star}$, $\delta_{i,k}^{\star}$, $\eta_{i,k}^{\star}$ denote the optimal transmission power, optimal Lagrange parameters for energy harvesting constraint, transmission power upper bound and lower bound constraint respectively. 
\par
From the above mentioned KKT conditions we get the expression for $p^{\star}_{i,k}$ by following the same approach proposed in \cite{ref3}. The analytic derivation for the optimal sensing power is mentioned below.
\par
\textit{Lemma 1}: Let $i$ and $j$ be any two arbitrary users, $i,j \in {1,2,\hdots,N}$, with $p_{i,k}^{\star}>0 $ and $p_{j,k}^{\star}=0$, then the following must hold:
\begin{equation}
\frac{c_{i,k}}{\lambda d_{i,k}+e_{i,k}} \geq \frac{c_{j,k}}{\lambda d_{j,k}+e_{j,k}}, \hspace{2mm} \forall{k}
\end{equation}
where $c_{i,k},d_{i,k},e_{i,k}$ are given by the following expressions:
$c_{i,k}=\frac{T-\tau_{k}}{MT\log{2}}h_{i,k}a_{i,k}(1-\theta_{k}), 
d_{i,k}=\frac{T-\tau_{k}}{MT}g_{i,k}a_{i,k}(1-\theta_{k})$, and 
$e_{i,k}=\sum_{r=k}^{M}\mu_{i,r}^{\star}a_{i,r}(T-\tau_{r})(1-\theta_{r})$. 

\par
\begin{proof}
The proof is given in Appendix A. 
\end{proof}

Now let $\pi$  be a permutation over $\left\{1,2,\hdots,N \right\}$ such that $\frac{c_{\pi(i),k}}{\lambda d_{\pi(i),k}+e_{i,k}} \geq \frac{c_{\pi(j),k}}{\lambda d_{\pi(j),k}+e_{j,k}}$ when $i<j$, $i,j \in \left\{1,2,\hdots,N\right\}$. Suppose there are 
$|\mathcal{I}|$ users that can transmit, where $\mathcal{I} \subseteq \left\{1,2,\hdots,N\right\}$ denotes this set of users. It can be verified that $\mathcal{I}=\left\{\pi(1),\hdots,\pi(|\mathcal{I}|)\right\}$. The following Lemma  provides a way to determine the cardinality of the set $\mathcal{I}$.

\textit{Lemma 2}: The cardinality of the set $\mathcal{I}$, $|\mathcal{I}|$ is the largest value of $x$ such that
\begingroup
\setlength\abovedisplayskip{1pt}
\setlength\belowdisplayskip{1pt}
\setlength{\arraycolsep}{-0.1em}
\begin{flalign}
& \frac{c_{\pi(x),k}}{\lambda d_{\pi(x),k}+e_{\pi(x),k}} >  \nonumber \\
& 1+ \sum_{b=1}^{x-1}P_{max}h_{\pi(b),k}a_{\pi(b),k}(1-\theta_{k}) 
\end{flalign}
\endgroup
and the optimal transmission power for $\pi(|\mathcal{I}|)^{th}$ user is given by the following expression:
\begingroup
\setlength\abovedisplayskip{1pt}
\setlength\belowdisplayskip{1pt}
\setlength{\arraycolsep}{-0.1em}
\begin{eqnarray}\label{eq:e8}
& p_{\pi(|\mathcal{I}|),k}^{\star}=\min(P_{max},(\frac{c_{\pi(|\mathit{I}|),k}}{\lambda d_{\pi(|\mathit{I}|),k}+e_{\pi(|\mathcal{I}|),k}} \nonumber \\
& -1-\sum_{b=1}^{|\mathcal{I}|-1}P_{max}h_{\pi(b),k}a_{\pi(b),k})\frac{1}{h_{\pi(|\mathcal{I}|),k}} \nonumber \\
\end{eqnarray}
\endgroup

\begin{proof}
The proof is given in Appendix B. 
\end{proof}

Note that the solution for $p^{\star}_{\pi(c),k}$ is $P_{max}$ if $c <|\mathcal{I}|$. If $c >|\mathcal{I}|$ , then $p^{\star}_{\pi(c),k}=0$. Only for the case when $c=|\mathcal{I}|$, then $p^{\star}_{\pi(c),k}$ is given by (\ref{eq:e8}).
\par
As an example if we consider a specific case of  $M=2$ and number of SUs $N=3$, then for any horizon if $\tau_{k}$ is fixed, the optimal transmission power would lie in the open set $(0,P_{max})$ for at most one of the 3 SUs. All the other SUs either transmit with $P_{max}$ or do not transmit.

\subsection{Optimal Sensing Time Policy}
Fixing the transmission power would result in the optimization problem in $\{\tau_k\}$:
\begingroup
\setlength\abovedisplayskip{1pt}
\setlength\belowdisplayskip{1pt}
\setlength{\arraycolsep}{-0.1em}
\begin{eqnarray}
& \underset{\{\tau_{k}\}}{\text{max}} \hspace{2mm}
& \sum_{k=1}^{M}\frac{T-\tau_{k}}{MT}\log_{2}(1+\sum_{i=1}^{N}p_{i,k}h_{i,k}a_{i,k}(1-\theta_{k}))  \nonumber\\
& \text{s.t.} \hspace{2mm}
&  \frac{1}{M}\sum_{k=1}^{M}\frac{T-\tau_{k}}{T}\sum_{i=1}^{N}p_{i,k}g_{i,k}a_{i,k}(1-\theta_{k}) \leq Q_{avg} \nonumber \\
& & 0 \leq \tau_{k} \leq T;  1 \leq k \leq M \nonumber \\
& & \sum_{r=1}^{k}a_{i,r}(p_s\tau_{r}+p_{i,r}(T-\tau_{r})(1-\theta_{r})) \leq B_{i}+\sum_{r=1}^{k-1}H_{i,r} \nonumber
\end{eqnarray}
\endgroup
The optimization problem can be simplified as follows:
\begingroup
\setlength\abovedisplayskip{1pt}
\setlength\belowdisplayskip{1pt}
\setlength{\arraycolsep}{-0.1em}
\begin{eqnarray}\label{eq:a15}
& \underset{\mathbf{\tau}}{\text{min}}
&\hspace{2mm} \mathbf{s}^{T}\boldsymbol{\tau} \nonumber\\
& \text{s.t.}
&  \hspace{2mm}\mathbf{w}^{T}\boldsymbol{\tau} \geq \tilde{Q}, \;
 \mathbf{0} \preceq \boldsymbol{\tau}\preceq T. \mathbf{1}, \;
 \mathbf{Y}^{T}\boldsymbol{\tau} \succeq \mathbf{z} 
\end{eqnarray}
\endgroup
where $\mathbf{s},\boldsymbol{\tau},\mathbf{w},\tilde{Q},\mathbf{Y},\mathbf{z}$ are given by the following expressions
\begingroup
\setlength\abovedisplayskip{1pt}
\setlength\belowdisplayskip{1pt}
\setlength{\arraycolsep}{-0.1em}
\begin{eqnarray*}
& & s_{k}= \log_{2}(1+\sum_{i=1}^{N}p_{i,k}h_{i,k}a_{i,k}(1-\theta_{k})) \; \nonumber \\
& & \mathbf{s}=[s_{1},s_{2},\hdots,s_{M}]^{T}, \; \boldsymbol{\tau}=[\tau_{1},\tau_{2},\hdots,\tau_{M}] ^{T}\nonumber \\
& & w_{k}=\sum_{i=1}^{N}p_{i,k}g_{i,k}a_{i,k}(1-\theta_{k}), \; \mathbf{w}=[w_{1},w_{2},\hdots,w_{M}]^{T} 
\nonumber \\
& & \tilde{Q}=T ((\sum_{k=1}^{M}w_{k})-MQ_{avg}), \; \rho_{i,r}=a_{i,r}\left\{p_{i,r}(1-\theta_{r})-p_s\right\} \nonumber \\
& & \mathbf{\Gamma}_{i}=\begin{bmatrix}
\rho_{i,1} & 0 & 0  & 0\\ 
\rho_{i,1} & \rho_{i,2} & 0  & 0 \\ 
\vdots &  \vdots & \ddots  & 0 \\ 
\rho_{i,1} & \rho_{i,2}  &  \hdots  & \rho_{i,M} 
\end{bmatrix}, \; \mathbf{Y}=[\mathbf{\Gamma}_{1}, \mathbf{\Gamma}_{2}, \hdots, \mathbf{\Gamma}_{N}]^{T}
 \end{eqnarray*}
\endgroup
\begingroup
\setlength\abovedisplayskip{1pt}
\setlength\belowdisplayskip{1pt}
\setlength{\arraycolsep}{-0.1em}
\begin{eqnarray*}
& & z_{i,k}=\sum_{r=1}^{k}a_{i,r}p_{i,r}T(1-\theta_{r})-(B_{i}+\sum_{r=1}^{k-1}H_{i,r}) \nonumber \\
& & \mathbf{z}=[z_{1,1}, \hdots, z_{1,M}, \hdots,  z_{N,1},\hdots, z_{N,M}]^{T}
\end{eqnarray*}
\endgroup
and $\mathbf{0},\mathbf{1}$ are $M$ dimensional vector of all $0$'s and $1$'s respectively. Problem (\ref{eq:a15}) is a linear program, which can be solved using a standard LP solver  with a reasonable accuracy and complexity.
\par
\textbf{Convergence analysis of alternating convex optimization}: The objective function of the optimization problem in (\ref{eq:e9}) is monotonically increasing  during the alternating convex optimization  procedure as at each iteration it is maximized with respect to a parameter by fixing the other parameter. Hence the value of the objective function either increases or remains unchanged. But it is important to note that even by considering only the short term constraints (\ref{eq:e1}) and (\ref{eq:e2}), and removing the average interference constraint in an extreme scenario, i.e, $a_{i,k}=1 \hspace{2mm} ,\forall{i,k}, \theta_{k}=0 \hspace{2mm} ,\forall{k}$, $\tau^{\star}_{k}=0 \hspace{2mm} ,\forall k$ and $p^{\star}_{i,k}=P_{max} \hspace{2mm} ,\forall{i,k}$, the sum capacity would be upper bounded by the expression $ \mathrm{E} \left\{\sum_{k=1}^{M}\frac{1}{M}\log_{2}(1+\sum_{i=1}^{N}P_{max}h_{i,k})\right\}$, where the expectation is taken over $h_{i,k}$. Using Jensen's inequality, we can write:
\begin{eqnarray}
\mathrm{E} \left\{\sum_{k=1}^{M}\frac{1}{M}\log_{2}(1+\sum_{i=1}^{N}P_{max}h_{i,k})\right\} \nonumber \\
\leq \sum_{k=1}^{M}\frac{1}{M}\log_{2}(1+\sum_{i=1}^{N}P_{max} \mathrm{E} \left\{h_{i,k}\right\}) \nonumber \\
=\sum_{k=1}^{M}\frac{1}{M}\log_{2}(1+\sum_{i=1}^{N}P_{max}) \nonumber 
\end{eqnarray}
Thus the objective function  is upper bounded by the sum capacity achievable by a feasible solution  for $a_{i,k},\theta_{k}, p_{i,k},\tau_{k}$ for a relaxed constraint set, which indicates that a upper bounded monotonically non-decreasing function must converge. Since the original problem is non-convex, this convergence is only guaranteed to reach a local optimum.

\section{Causal Optimization with Finite Battery}
While the non-causal information pattern is unrealistic, it serves as a benchmark for the more realistic scenario of causal CSI and battery state information with finite battery capacity. The resulting problem can be formulated as a finite horizon stochastic control problem and thus solved by a dynamic programming algorithm.  The values of $a_{i,k}$ and $\theta_{k}$, we can be found either via an exhaustive search or suboptimally by the previously discussed heuristic policy depending on the complexity concerned with a given scenario. 
\subsection{Information Pattern}
During each time slot $k$, FC receives the CSI between SU transmitter and PU receiver $\mathbf{g}_{k}=\left\{g_{1,k},g_{2,k},\hdots,g_{N,k}\right\}$ causally, either via cooperation from the PU receiver (or base station), or via feedback from a cooperative node located close to the PU receiver. The CSI between SU transmitter and FC $\mathbf{h}_{k}=\left\{h_{1,k},h_{2,k},\hdots,h_{N,k}\right\}$ and the SU battery state information $\mathbf{B}_{k}=\left\{B_{1,k},B_{2,k},\hdots,B_{N,k}\right\}$ are assumed to be received by FC via typical channel estimation techniques and feedback from the SUs to the FC. Information available to FC at $k^{th}$ horizon is given by the tuple $\mathbf{J}_{k}=\left\{\mathbf{g}_{k},\mathbf{h}_{k},\mathbf{B}_{k},\mathbf{J}_{k-1}\right\}$. 
\subsection{Dynamic Programming Algorithm}
Now we discuss the process of finding the optimal \textit{sensing time} and \textit{transmission power} for energy harvesting SUs with finite battery to maximize the sum-capacity at FC under the assumption that the information about channel gain and energy arrival process of all SUs are causally available at the FC.
We assume the sum-capacity expression corresponding to a fixed horizon $k$ can be written as
\begin{eqnarray}
C(p_{i,k},\tau_{k})=\frac{T-\tau_{k}}{MT}\log_{2}(1+\sum_{i=1}^{N}p_{i,k}h_{i,k}a_{i,k}(1-\theta_{k})) \nonumber \\
-\lambda (\frac{T-\tau_{k}}{MT}\sum_{i=1}^{N}p_{i,k}g_{i,k}a_{i,k}(1-\theta_{k})-Q_{avg}) \label{eq:a12}
\end{eqnarray}
Here $\lambda$ is the Lagrange parameter corresponding to average interference constraint.
\par
 We first define the feasible set for the optimization variables as:
\begin{equation}
{S}=\left\{(p_{i,k},\tau_{k}): p_{i,k},\tau_{k} \hspace{2mm} \text{satisfy} (\ref{eq:e1}),(\ref{eq:e2}),(\ref{eq:e3})\right\} \nonumber
\end{equation}
With $\lambda$  fixed, the optimal value of \textit{transmission power} and \textit{sensing time} can be determined by the following theorem:
\newtheorem{thm}{Theorem}
\begin{thm}
 With the initial condition  $\mathbf{J}_{1}=\left\{\mathbf{g}_{1},\mathbf{h}_{1},\mathbf{B}_{1}\right\}$,  the value of the finite horizon finite battery problem with causal information is given by $V_{1}(\mathbf{g}_{1},\mathbf{h}_{1},\mathbf{B}_{1})$, which can be computed by the backward Bellman dynamic programming equation:
\begin{eqnarray}
& V_{k}(\mathbf{g},\mathbf{h},\mathbf{B})=\max_{(p_{i,k},\tau_{k}) \in \mathcal{S}}[C(p_{i,k},\tau_{k})+  \nonumber \\
& \mathrm{E}[V_{k+1}(\mathbf{g}_{k+1},\mathbf{h}_{k+1},\mathbf{B}_{k+1}|p_{i,k},\tau_{k})]] \nonumber
\end{eqnarray}
\end{thm}
\par
\begin{proof}
The proof can be obtained by the optimality conditions for the finite horizon stochastic control problem \cite{ref13}.
\end{proof}
The solution of the causal optimization problem, which can be computed numerically by searching over discretized values of the optimization variables, is obtained as:
\begin{eqnarray}
& \left\{p_{i,k}^{\star},\tau_{k}^{\star}\right\}=\text{argmax}_{p_{i,k},\tau_{k} \in \mathcal{S}}[C(p_{i,k},\tau_{k}) + \nonumber \\
& \mathbf{E}[V_{k+1}(\mathbf{g}_{k+1},\mathbf{h}_{k+1},\mathbf{B}_{k+1}|p_{i,k},\tau_{k})]]  \nonumber
\label{eq:bellmansol}
\end{eqnarray}
The optimal value of $\lambda$ from (\ref{eq:a12}) is found by the solving the following equation:
\begin{equation}
\lambda \left\{\mathrm{E} \left\{\sum_{k=1}^{M}\frac{T-\tau_{k}}{T}\sum_{i=1}^{N}p_{i,k}g_{i,k}a_{i,k}(1-\theta_{k})\right\}-Q_{avg}\right\}=0 \label{eq:e5}
\end{equation}
We solve (\ref{eq:e5}) using a bisection algorithm, in which
\begin{itemize}
\item The variables $p_{i,k}^{\lambda}$ and $\tau_k^{\lambda}$ are obtained  based on a fixed $\lambda$ from 
(\ref{eq:bellmansol}). Hence, we express the interference term corresponding to fixed $\lambda$ as $G(\lambda)=\mathrm{E}\left\{\sum_{k=1}^{M}\frac{T-\tau_k^{\lambda}}{T}\sum_{i=1}^{N}{p_{i,k}^{\lambda}} g_{i,k}a_{i,k}(1-\theta_{k})\right\}-Q_{avg}$.
\item The parameter $\lambda^{(k)}$ denotes the $k^{th}$ iteration of $\lambda$ in the Bisection iterative procedure.
\item Two initial points $\lambda_{1}^{(1)}$ and $\lambda_{2}^{(1)}$ are chosen from the feasible set $\Lambda=\left\{\lambda: \lambda>0\right\}$, such that $\lambda_{1}^{(1)} G(\lambda_{1}^{(1)}) <0$ and $\lambda_{2}^{(1)} G(\lambda_{2}^{(1)}) >0$.
\item The bisection method is continued until $|\lambda^{(k)}G(\lambda^{(k)})|<\epsilon$, where $\epsilon>0$ is a predefined threshold of convergence.
\item The Bisection method converges since the expression $G(\lambda)$  can be shown to be monotonically decreasing with  $\lambda$.
\end{itemize}
 Note that this procedure is performed offline purely based on the statistics of the channel gains and harvested energy information.
Based on this procedure, the FC creates a lookup table for optimal values of $p_{i,k}$ and $\tau_{k}$ corresponding to discrete quantized values of $\mathbf{g}_{k},\mathbf{h}_{k},\mathbf{H}_{k}$. In real time FC receives the channel gains and battery states and check for closest quantization point in its lookup table. The optimal sensing and transmission power are fetched from the look-up table and  sent to the individual SUs (via errorfree feedback links with negligible delay), which are then used by the SUs for sensing and information transmission.
%

\section{Simulation Results}

\begin{figure}[t]
\centering
\includegraphics[scale=0.50]{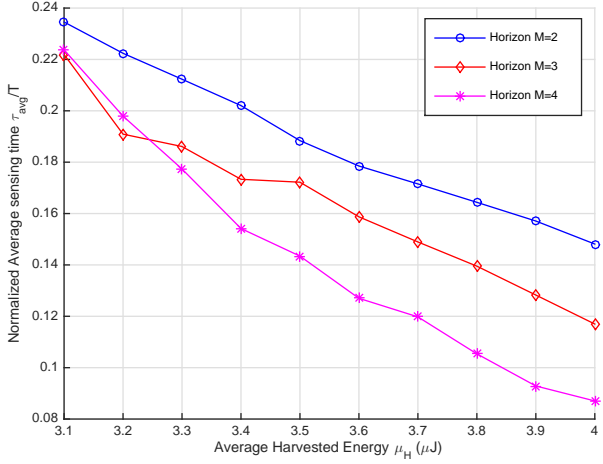}
\caption{Normalized average sensing time $\tau_{avg}$ vs Average harvested energy $\mu_{H}$  with ratio of SU Tx-FC Rx and SU Tx-PU Rx average channel gain $\mu_{h}/\mu_{g}$ fixed.}
\label{fig:fig1}
\end{figure}

\begin{figure}[t]
\centering
\includegraphics[scale=0.45]{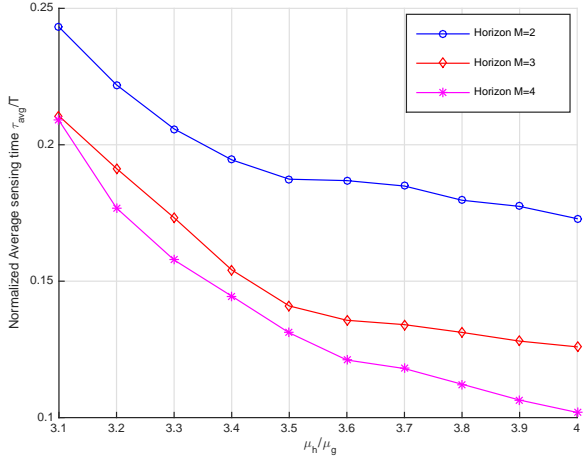}
\caption{$\tau_{avg}$ vs ratio of SU Tx-FC Rx and SU Tx-PU Rx average channel gain $\mu_{h}/\mu_{g}$ with average harvested energy $\mu_{H}$ fixed.}
\label{fig:fig2}
\end{figure}

\begin{figure}[t]
\centering
\includegraphics[scale=0.50]{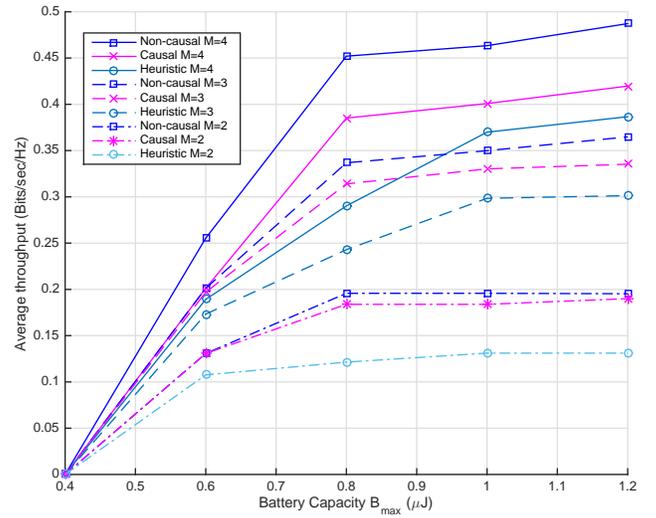}
\caption{Average throughput vs $B_{max}$  with causal and non-causal information pattern with $M=2,3,4$.}
\label{fig:fig3}
\end{figure}

\begin{figure}[t]
\centering
\includegraphics[scale=0.50]{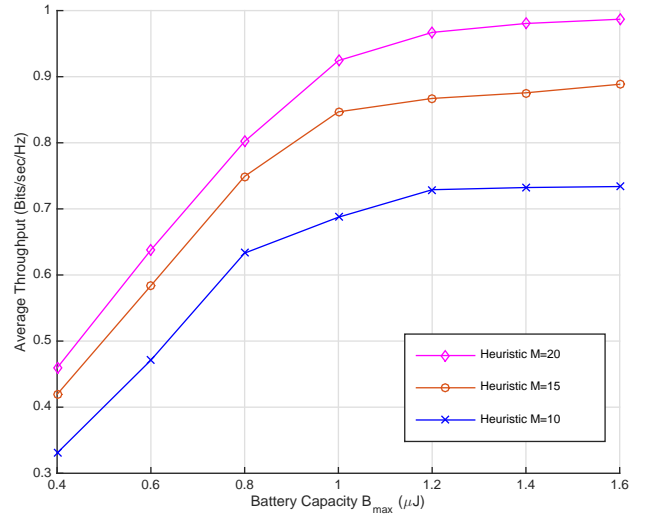}
\caption{Average throughput vs $B_{max}$  for the heuristic policy with $M=10,15,20$.}
\label{fig:fig4}
\end{figure}

In this section we present numerical results for the causal and non-causal CSI and battery state scenarios for the optimization problem under consideration. The channels $g_{i,k},h_{i,k}$ are modeled as exponentially distributed random variables with unity mean unless otherwise stated. We assume the energy harvesting process is also an exponentially distributed random process with unity mean. The PU activity probability is set to $\kappa=0.8$. The sensing channel signal to noise ratio (SNR) is assumed to be $-15$dB. The PU signal variance is taken to be $\sigma^{2}_{s}=1$ mW. In all simulation studies, we assume $N=2$ SUs. The length of a time slot is taken as $T=2$ ms. The probability of false alarm $P_{fa}$ is taken to be $0.03$. The sampling frequency is assumed to be 1 MHz and the normalized threshold of detection is assumed to be $\frac{\gamma}{\sigma^{2}_{n}}=1.006$. This corresponds to  a minimum sensing time limit $\tau_{l}= 0.1$ ms. For a fair comparison, this constraint on the minimum sensing time has been applied to all of the non-causal, causal and heuristic policy based methods. Transmission power is assumed to be upper bounded by $P_{max}=1$ mW. The initial battery level for each SU is assumed to be $0.4$ mW. For dynamic programming $\mathbf{g}_{k},\mathbf{h}_{k},\mathbf{H}_{k}$ are quantized into 5 different discrete levels. For the first two plots we vary the length of horizon among $M=2,3,4$. In Fig.\ref{fig:fig1}, we plot the average sensing time for non-causal CSI and battery state scenario denoted by $\tau_{avg}$ with respect to average harvested energy denoted by $\mu_{H}$, keeping the mean of the channel distribution denoted by $\mu_{g}$ and $\mu_{h}$ constant. The averages are taken over $50$ Monte-Carlo simulations. From Fig.\ref{fig:fig1}, it is evident that average sensing time $\tau_{avg}$ decreases monotonically by increasing the number of horizons. This is due to the fact that increasing the length of horizon $M$ in the non-causal CSI and battery state scenario helps to spread out the sensing time over multiple time slots, which means on average transmission time increases and sensing time decreases with increasing $M$. Fig.\ref{fig:fig2} demonstrates of average sensing time $\tau_{avg}$ with respect to the ratio of channel mean $\mu_{h}/\mu_{g}$ keeping $\mu_{H}$ fixed. The nature of the plot is similar to the Fig.\ref{fig:fig1} as better SU direct channel gains compared to interference channel gains provide favourable transmission conditions which also increase transmission time and reduce sensing time.  Fig.\ref{fig:fig3} shows the optimized average throughput, i.e. average sum-capacity plotted against battery capacity $B_{max}$ for non-causal CSI and battery state scenario with exhaustive search technique and causal CSI and battery state scenario with exhaustive search and heuristic policy. As expected the average throughput increases with increasing horizon and non-causal CSI and battery state scenario provides an upper bound for the causal counterpart. Also the heuristic policy by choosing $a_{i,k}$ and $\theta_{k}$ performs inferior to both of them, but serves as a less computationally complex alternative to its optimal counterpart. As a numerical comparison, it should be noted that for $M=3$ and $B_{max}=1 \mu J$ the average throughput loss with exhaustive search and heuristic policy in causal CSI and battery state scenario compared to non-causal counterpart are $5.7$ and $14.7$ percent respectively. Fig.\ref{fig:fig4} shows the heuristic policy-based optimized average throughput as a function of varying battery capacity in the causal CSI and battery state scenario with finite battery. These plots are obtained for three different horizons $M=10,15,20$, significantly longer horizon lengths due to the reduced complexity of the heuristic policy.

\section{Conclusion}
This paper investigated a MINLP problem of maximizing the expected achievable  sum throughput in a fading multiple access CR network with the energy harvesting constraints where the CR nodes perform cooperative spectrum sensing. An analytical solution  is obtained for the system with non-causal CSI and infinite battery capacity. This provides an upper bound on the throughput of  a more realistic scenario involving causal CSI and finite battery capacity, which can be formulated as a stochastic control problem and solved using a dynamic programming algorithm. To combat the exponential complexity of handling exhaustive search policies involving the decision variables in the causal CSI scenario, a heuristic policy is proposed. The problem can be extended to incorporate the concept of infinite horizon optimization and energy sharing between the SUs, and also the innovation of better performing heuristic policies.

\appendices
\section{Proof of Lemma 1}
\begin{proof} 
Since $p_{j,k}^{\star}=0$ and $p_{i,k}^{\star}> 0$ from the KKT condition it follows that $\delta_{j,k}^{\star}=0$ and $\eta_{i,k}^{\star}=0$. So it can be deduced that:
\begin{align}
\frac{\frac{T-\tau_{k}}{MT}h_{i,k}a_{i,k}(1-\theta_{k})}{1+\sum_{l=1}^{N}p_{l,k}^{\star}h_{l,k}a_{l,k}(1-\theta_{k})}-\frac{T-\tau_{k}}{MT}\lambda g_{i,k}a_{i,k}(1-\theta_{k})  \nonumber \\
-\sum_{r=k}^{M} \mu_{i,r}^{\star}(T-\tau_{r})a_{i,r}(1-\theta_{r}) \geq 0 
\end{align}
\begin{align}
\frac{\frac{T-\tau_{k}}{MT}h_{j,k}a_{j,k}(1-\theta_{k})}{1+\sum_{l=1}^{N}p_{l,k}^{\star}h_{l,k}a_{l,k}(1-\theta_{k})}-\frac{T-\tau_{k}}{MT}\lambda g_{j,k}a_{j,k}(1-\theta_{k})  \nonumber \\
-\sum_{r=k}^{M} \mu_{j,r}^{\star}(T-\tau_{r})a_{j,r}(1-\theta_{r}) \leq 0 
\end{align}
From the above two equation the given lemma can be easily obtained.
\end{proof}

\section{Proof of Lemma 2}
\begin{proof}
To prove it we have need the following lemma:

\textit{Lemma 3}: The optimal solution of the problem has at most one user indexed by $i$ that satisfies $0<p_{i,k}^{\star} <P_{max}$ where $i=\pi(|\mathcal{I}|)$, and the following condition must hold for the optimal transmission power:
\begin{eqnarray}
& \sum_{c=1}^{|\mathcal{I}|}p_{\pi(c),k}^{\star}h_{\pi(c),k}a_{\pi(c),k}(1-\theta_{k}) = \nonumber \\
& \frac{c_{\pi(|\mathcal{I}|),k}}{\lambda d_{\pi(|\mathcal{I}|),k}+e_{\pi(|\mathcal{I}|),k}}-1 
\end{eqnarray}
\begin{proof}
(By contradiction) We assume that there exist two users $i$ and $j$ such that with $0<p_{i,k}^{\star}<P_{max}$ and $0< p_{j,k}^{\star}<P_{max}$. From the KKT condition we determine that $\delta_{i,k}^{\star}=\delta_{j,k}^{\star}=0$ and $\eta_{i,k}^{\star}=\eta_{j,k}^{\star}=0$ respectively. Using this values we can write:
\begin{equation}
\frac{c_{i,k}}{\lambda d_{i,k}+e_{i,k}}= \frac{c_{j,k}}{\lambda d_{j,k}+e_{j,k}} 
\end{equation}
Since $h_{i,k}$ and $g_{i,k}$'s are independent of $h_{j,k}$ and $g_{j,k}$'s and they are drawn from a continuous distribution and $\lambda$ is constant, it can be inferred that the above equality is satisfied with a probability zero. Thus we can deduce that there is at most one user $i$ with $0<p_{i,k}^{\star}<P_{max}$. The optimal transmission power expression for the user $i$ mentioned in the Lemma can be proven as the following:
\par
Using the KKT condition and the expression of the proof it is easy to see that for any user $f \in \mathcal{I}, f \neq i$, with $p_{f,k}^{\star}>0$ must satisfy
\begin{equation}
\frac{c_{f,k}}{\lambda d_{f,k}+e_{f,k}} \geq \frac{c_{f,k}}{\lambda d_{f,k}+e_{f,k}} 
\end{equation}
Thus we can say that $i=\pi(\mathcal{|I|})$.
\par
From the previous Lemma, we can infer that there are only two possible sets of solutions for $p_{i,k}^{\star},k \in \mathcal{I}$:
\begin{itemize}
\item Case I: $p_{\pi(a),k}^{\star}=P_{max},a=1,2,\hdots,|\mathcal{I}|$.
\item Case II:  $p_{\pi(a),k}^{\star}=P_{max},a=1,2,\hdots,|\mathcal{I}|-1$
\end{itemize}
The expression for $p^{\star}_{\pi(|\mathcal{I}|),k}$ is only relevant when $a_{i,k}=1$ and $\theta_{k}=0$, which can be written as :
\begin{flalign}
& p^{\star}_{\pi(|\mathcal{I}|),k}= (\frac{c_{\pi(|\mathcal{I}|),k}}{\lambda d_{\pi(|\mathcal{I}|),k}+e_{\pi(|\mathcal{I}|),k}}-1 \nonumber \\
& -\sum_{b=1}^{|\mathcal{I}|-1}P_{max}h_{\pi(b),k}a_{\pi(b),k}) \frac{1}{h_{\pi(|\mathcal{I}|),k)}} 
\end{flalign}
Since $p_{\pi(|\mathcal{I}|),k} ^{\star}\leq P_{max}$, thus we can write:
\begin{eqnarray}\label{eq:eq2}
& p_{\pi(|\mathcal{I}|),k}^{\star}=\min(P_{max},(\frac{c_{\pi(|\mathit{I}|),k}}{\lambda d_{\pi(|\mathit{I}|),k}+e_{\pi(|\mathcal{I}|),k}} \nonumber \\
& -1-\sum_{b=1}^{|\mathcal{I}|-1}P_{max}h_{\pi(b),k}a_{\pi(b),k})\frac{1}{h_{\pi(|\mathcal{I}|),k}} \nonumber \\
\end{eqnarray}
The remaining part is showing the fact that optimal number of active users $|\mathcal{I}|$ is the largest value of $x$ such that :
\begin{flalign}
\frac{c_{\pi(x),k}}{\lambda d_{\pi(x),k}+e_{\pi(x),k}} > 1+ \nonumber \\
& \sum_{b=1}^{x-1}P_{max}h_{\pi(b),k}a_{\pi(b),k}(1-\theta_{k}) 
\end{flalign}
It can be shown that both the case I and case II, any SU $\pi(b), b=1,\hdots,|\mathcal{I}|$, the above inequality holds. Since from Lemma 1, it can be said that its left hand side decreases with $x$ while right hand side increases with $x$, thus it is sufficient to show that the inequality holds for $b=|\mathcal{I}|$. Thus in this case $\delta^{\star}_{\pi({|\mathcal{I}|}),k} \geq 0$ and $\eta^{\star}_{\pi({|\mathcal{I}|}),k}=0$, we have
\begin{flalign}
&\frac{c_{\pi(|\mathcal{I}|),k}}{\lambda d_{\pi(|\mathcal{I}|),k}+e_{\pi(|\mathcal{I}|),k}}  \nonumber \\
&  \geq  1+ \sum_{b=1}^{|\mathcal{I}|}p^{\star}_{\pi(b),k}h_{\pi(b),k}a_{\pi(b),k}(1-\theta_{k}) \nonumber \\
& > \sum_{b=1}^{|\mathcal{I}|-1}P_{max}h_{\pi(b),k}a_{\pi(b),k}(1-\theta_{k}) 
\end{flalign}

Next we have to show that for any user $\pi(j), j=\mathcal{|I|}+1,\hdots,M$, the inequality does not hold. Again it is sufficient to show that it does not hold for $\pi(|\mathcal{I}|+1)$. For that user  $\delta^{\star}_{\pi({|\mathcal{I}|+1}),k}= 0$ and $\eta^{\star}_{\pi({|\mathcal{I}|+1}),k} \geq 0$, we have
\begin{flalign}
& \frac{c_{\pi(|\mathcal{I}|+1),k}}{\lambda d_{\pi(|\mathcal{I}|+1),k}+e_{\pi(|\mathcal{I}|+1),k}}  \nonumber \\
& \leq  1+ \sum_{b=1}^{|\mathcal{I}|}p^{\star}_{\pi(b),k}h_{\pi(b),k}a_{\pi(b),k}(1-\theta_{k}) \nonumber \\
& \leq \sum_{b=1}^{|\mathcal{I}|}P_{max}h_{\pi(b),k}a_{\pi(b),k}(1-\theta_{k}) 
\end{flalign}
Thus we can determine $|\mathcal{I}|$. 
\end{proof}
\end{proof}

\section{Proof of Lemma 4}
\begin{proof}
This 2-horizon problem can be simplified as
\begin{eqnarray}
& \underset{\tau_{1},\tau_{2}}{\text{min}}
&  \hspace{2mm}s_{1}\tau_{1}+s_{2}\tau_{2} \nonumber \\
& \text{s.t.} \hspace{2mm}
& w_{1}\tau_{1}+w_{2}\tau_{2} \geq \tilde{Q} \nonumber \\
& & \tau_{l} \leq \tau_{1},\tau_{2} \leq T  \nonumber \\
& & a_{i,1}(p_{s}\tau_{1}+p_{i,1}(T-\tau_{1})(1-\theta(1))) \leq B_{i}\; \forall{i} \nonumber \\
& & a_{i,1}(p_{s}\tau_{1}+p_{i,1}(T-\tau_{1})(1-\theta(1)))+ \nonumber \\
& & a_{i,2}(p_{s}\tau_{2}+p_{i,2}(T-\tau_{2})(1-\theta(2))) \leq B_{i}+H_{i,1} \; \forall{i} \nonumber
\end{eqnarray}
In the above equation the constants $s_{1},s_{2},w_{1},w_{2}$ represents the following:
\begin{eqnarray}
& & s_{k}=\log_{2}(1+\sum_{i=1}^{N}p_{i,k}h_{i,k}a_{i,k}(1-\theta_{k})), \hspace{2mm}k=1,2\nonumber \\
& &  w_{k}= \sum_{i=1}^{N}p_{i,k}g_{i,k}a_{i,k}(1-\theta_{k}), \hspace{2mm} k=1,2 \nonumber \\
& & \tilde{Q}=T(w_{1}+w_{2}-2Q_{avg}) \nonumber \\
\end{eqnarray}
In this scenario, if we assume that the sensing time for the first horizon $\tau_{1}$ is fixed, then the problem can be expressed as optimization over $\tau_{2}$. With this approach we have formulated the expression of $\tau_{2}$ in terms of $\tau_{1}$. 
\par

Considering the energy causality constraint for the second horizon we get a lower bound of $\tau_{2}$ in term of $\tau_{1}$ as follows:
\begin{eqnarray}
&& \tau_{2} \geq f_{i}(\tau_{1})= \nonumber \\
&& \frac{T(a_{i,1}p_{i,1}(1-\theta_{1})+a_{i,2}p_{i,2}(1-\theta_{2}))-B_{i}-H_{i,1}}{a_{i,2}(p_{i,2}(1-\theta_{2})-p_s)} \nonumber \\
&& -\frac{a_{i,1}(p_{i,1}(1-\theta_{1})-p_s)}{a_{i,2}(p_{i,2}(1-\theta_{2})-p_s)} \nonumber \; \forall{i}
\end{eqnarray}

Now as this inequality is valid for all users, we can write:
\begin{equation} \label{eq:eq3}
\tau_{2} \geq f({\tau_{1}})= \max \left \{f_{1}(\tau_{1}),\hdots,f_{N}(\tau_{1}) \right\} 
\end{equation}
 
Now another lower bound can be obtained form the interference constraint as the following:
\begin{equation}
\tau_{2} \geq g(\tau_{1})=\frac{\tilde{Q}-w_{1}\tau_{1}}{w_{2}} \nonumber
\end{equation}

The highest lower bound of $\tau_{2}$ is the optimal solution corresponding to the chosen fixed $\tau_{1}$, which can be written by the following expression:
\begin{equation}
\tau^{\star}_{2}=\max \left\{f(\tau_{1}),g(\tau_{1}),\tau_{l}\right\} \nonumber
\end{equation}
\end{proof}




\bibliographystyle{IEEEtran}
\renewcommand{\baselinestretch}{0.9}
\bibliography{IEEEfull,eusipco_ref}
%
%
%

\end{document}